\documentstyle[prl, aps, epsf]{revtex}

\begin{document}
\draft


\twocolumn[\hsize\textwidth\columnwidth\hsize\csname%
@twocolumnfalse\endcsname%

\title{From BCS to BEC Superconductivity: 
Spectroscopic Consequences}

\author{L. S. Borkowski$^*$ and C. A. R. S\'{a} de Melo}
\address{School of Physics, Georgia Institute of Technology, 
Atlanta, Georgia 30332}

\date{\today}

\maketitle

      \begin{abstract}

The evolution from BCS to BEC superconductivity in the s-wave and
d-wave channels is analyzed at zero temperature for 
a two-dimensional superconductor.
Spectroscopic quantities for s-wave and d-wave systems
present fundamental differences when particle density and
attraction strength are varied.
A detailed analysis of single quasiparticle properties
(excitation spectrum, momentum distribution, spectral function and density
of states) indicates that the evolution of these spectroscopic quantities
in the d-wave case is not smooth, unlike the situation encountered
for the s-wave system.
\end{abstract}
]

The problem of the evolution from BCS to BEC superconductivity is an
old one~\cite{leggett,NSR} but recently it has received considerable 
attention in connection with high temperature 
superconductors,~\cite{RDS,varma,zwerger,strinati,levin,adhikari,sdm}
where strong deviations from the BCS behavior have been observed
experimentally in spectroscopic quantities~\cite{expt}
at low temperatures. 
Furthermore, the recent discovery of BEC
in atomic systems~\cite{BEC}  and the exciting possibility
of BEC in atomic Fermi systems raises the
question of the evolution from BCS to BEC in atomic Fermi
systems as well.~\cite{stoof}

In this work we address the question of whether
the evolution of spectroscopic quantities from a BCS to a BEC 
superconductor is smooth at zero temperature. For this purpose we 
study the single quasiparticle properties (excitation spectrum, 
momentum distribution, spectral function and density of states) as
a function of attraction strength or particle density 
for the s-wave and d-wave cases. In anticipation of the main results, 
we must say that the evolution of spectroscopic properties in 
the s-wave case is smooth, while in the d-wave it is {\it not}.
The main reasoning for this statement is as follows. 

Quite generally the evolution from BCS to BEC superconductivity 
can be characterized by two parameters: the chemical potential
$\mu$ and the Cooper pair size $\xi_{pair}$. 
The BCS limit is characterized by a positive chemical potential 
${\mu = \epsilon_F}$
and a large size of Cooper pairs ${ (\xi_{pair} \gg k_F^{-1}) }$, 
while the BEC regime is characterized by a large and negative 
chemical potential ${ \mu = -E_{b}^{(\ell)} }$, 
where ${ E_{b}^{(\ell)}}$ is the binding energy of the 
two-body problem in the ${ \ell^{th}}$ angular momentum channel, 
and by a small size of pairs ${ (\xi_{pair} \ll k_F^{-1}) }$.
Here ${ \ell = 0}$ (or s) indicates the s-wave channel, 
while ${ \ell = 2}$ (or d) indicates the d-wave channel. 
The excitation spectrum at zero temperature has the form 
${ E_{\ell} ({\bf k}) 
= { \left ( (\epsilon_{\bf k} - \mu)^2 
+ |\Delta_{\ell} ({\bf k})|^2 \right )^{1/2}} }$, 
where ${ \epsilon_{\bf k} = k^2/2m}$ and
${ \Delta_{\ell} ({\bf k}) = \Delta_{0\ell}  h_{\ell} (k) 
\cos(\ell\phi)}$, with ${ k  = |{\bf k}| }$.
In the s-wave case the excitation spectrum ${ E_{s} ({\bf k})}$ 
is gapped for all $\bf k$, 
and it increases smoothly from the BCS to the BEC limit.
As a result the 
quantities that depend directly on the excitation spectrum 
${ E_{s} ({\bf k})}$ also evolve 
smoothly. For instance, 
the quasiparticle density of states $N_{s} (\omega)$ at low 
frequencies is always zero, since there are no available states inside 
the gap. Thus, contributions from single quasiparticle excitations 
to thermodynamic quantities  
are always exponentially small at low temperatures.
In the d-wave case the situation is qualitatively different. 
For ${\mu > 0}$ the 
superconductor is gapless at the Dirac points 
${ k = k_{\mu} = \sqrt{2m\mu} }$, 
${ \phi = \pm \pi/4, \pm 3\pi/4 }$, while for 
${ \mu < 0}$ the superconductor acquires a finite gap.
The line $\mu =0$ separates  
two regimes with qualitatively different behavior.
This has important consequences
for the momentum distribution, spectral function, and density of states. 
The quasiparticle density of states 
$N_d (\omega)$ changes 
discontinously at low 
frequencies from linear in ${ \omega}$ for ${ \mu > 0}$ 
(where ${ E_d ({\bf k}) }$ is linear in momentum close to the Dirac 
points), to a constant at ${ \mu = 0}$ (where ${ E_d ({\bf k}) }$ 
is quadratic for small momenta), to zero for 
${ \mu < 0}$ (where ${ E_d ({\bf k}) 
\simeq |\mu| + {\cal O} (k^2) }$
for small $k$). Thus, contributions from single quasiparticle 
excitations to thermodynamic quantities 
at low temperatures also exhibit singular
behavior in the vicinity of ${ \mu = 0}$. 

In order to analyze how the spectroscopic quantities change
from the BCS to BEC limit, 
we start with the two dimensional Hamiltonian
\begin{equation}
\label{eqn:hamiltonian}
H =
\sum_{{\bf k} \sigma} \epsilon_{\bf k} \psi^{\dagger}_{{\bf k} \sigma}
\psi_{{\bf k} \sigma}
+\sum_{{\bf k} {\bf k^{\prime}} {\bf q} } V_{\bf k k^{\prime}}
b^{\dagger}_{\bf k q} b_{\bf k^{\prime} q}
\end{equation}
where 
$ b_{\bf k q} = 
\psi_{{-\bf k} + {\bf q/2} \downarrow } 
\psi_{ {\bf k} + {\bf q/2} \uparrow }$.
The interaction potential ${ V_{\bf k k^{\prime}} }$ is expanded in 
its angular momentum components as 
$
{
V_{\bf k k^{\prime}} 
=  \sum_{\ell = - \infty}^{+ \infty} {V_{k k^{\prime}}^{(\ell)} 
\exp(i\ell \phi_{k k^\prime}),
}
}
$
where ${ \phi_{\bf k k^\prime} = {\rm acos} 
({\bf \hat k \cdot \hat k^{\prime} }) }$ 
is the angle betwen the vectors ${ \bf k}$ and ${ \bf  k^{\prime}}$ and
${
V_{k k^{\prime}}^{(\ell)} 
= 2\pi \int_0^{\infty} dr r J_{\ell} (kr) J_{\ell} (k^\prime r) V(r). 
}
$
The index ${ \ell}$ labels angular momentum states in two spatial 
dimensions, with $\ell = 0, \pm 1, \pm 2,...$, corresponding
to s,p,d,..., channels respectively. 
A possible choice of the real space potential can be
${ V(r) = V_1 \Theta (R_1 - r) - V_0 \Theta (r - R_1) \Theta (R_0 - r)},$
which is repulsive at short distances ${ r < R_1}$, attractive 
at intermediate distances ${R_1 < r < R_0}$, and vanishes for ${ r > R_0}$.
Quite generally any short ranged real space potential ${ V(r)}$ 
with range ${ R_0}$ leads to a ${ V_{k k^{\prime}}^{(\ell)}}$ 
which is separable for small momenta, provided that 
${ kR_0 \ll 1}$ {\it or} ${ k^\prime R_0 \ll 1}$. 
In the simpler limit when both
${ kR_0 \ll 1}$ {\it and} ${ k^\prime R_0 \ll 1}$,
$
{ V_{k k^{\prime}}^{(\ell)} \simeq k^\ell k^{\prime \ell} 
{(2\pi / 2^{2\ell}) } \int_0^\infty dr r^{2\ell + 1} V(r) 
},
$
where ${ \ell}$ is assumed to be positive for definiteness.
Notice here that ${ V_{k k^{\prime}}^{(\ell)} \propto k^\ell 
k^{\prime \ell}}$, thus for the s-wave 
case ${ V_{k k^{\prime}}^{(s)} \propto \mbox{const.} }$, 
while for the d-wave case
${ V_{k k^{\prime}}^{(d)} \propto k^2 k^{{\prime}2} }$. 
In the opposite limit, ${ kR_0 \gg 1}$ {\it or} ${k^\prime R_0 \gg 1}$, 
the potential ${ V_{k k^{\prime}}^{(\ell)} }$ is certainly not 
separable.
In the simpler limit when both 
${ kR_0 \gg 1}$ {\it and} ${ k^\prime R_0 \gg 1}$, 
${ V_{k k^{\prime}}^{(\ell)} }$  
mixes different ${ \bf k}$ and ${ \bf k^{\prime} }$ and
shows an oscillatory behavior which is dependent on the exact form of 
${ V(r)}$, with a decaying envelope proportional to 
${ k^{-1/2} {k^{\prime}}^{-1/2} }$. 

\begin{figure}
\begin{center}
\hskip 1.5cm
\epsfxsize=4.7cm
\epsffile{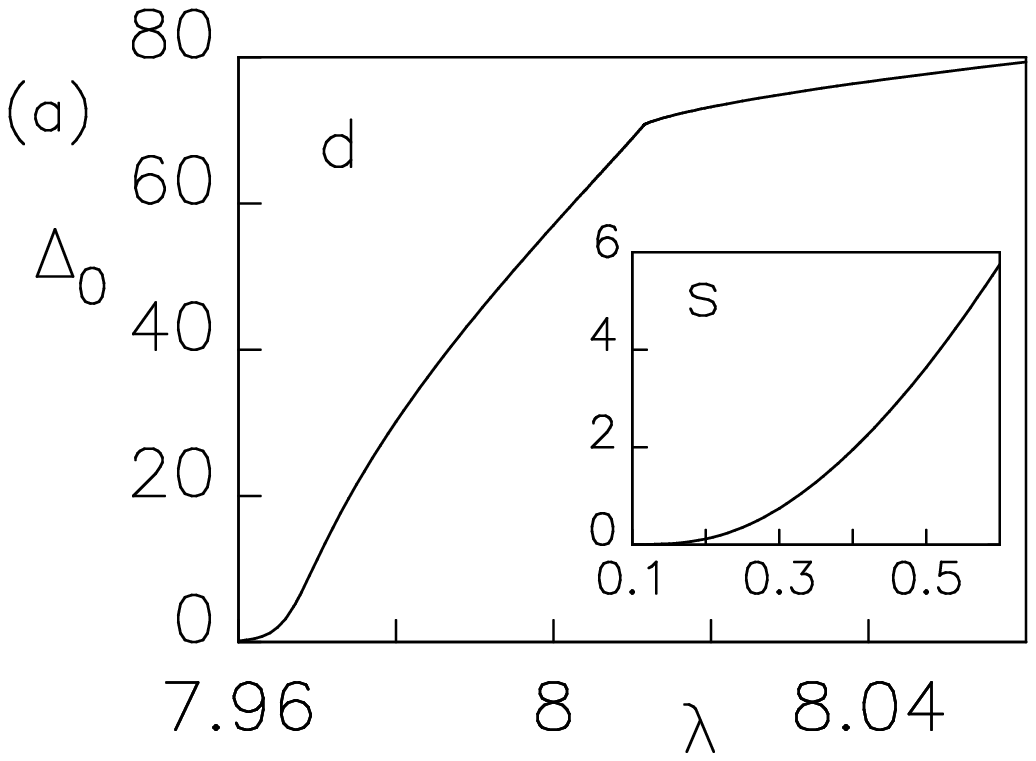}
\vskip -1.5cm
\hskip 1.5cm
\epsfxsize=4.7cm
\epsffile{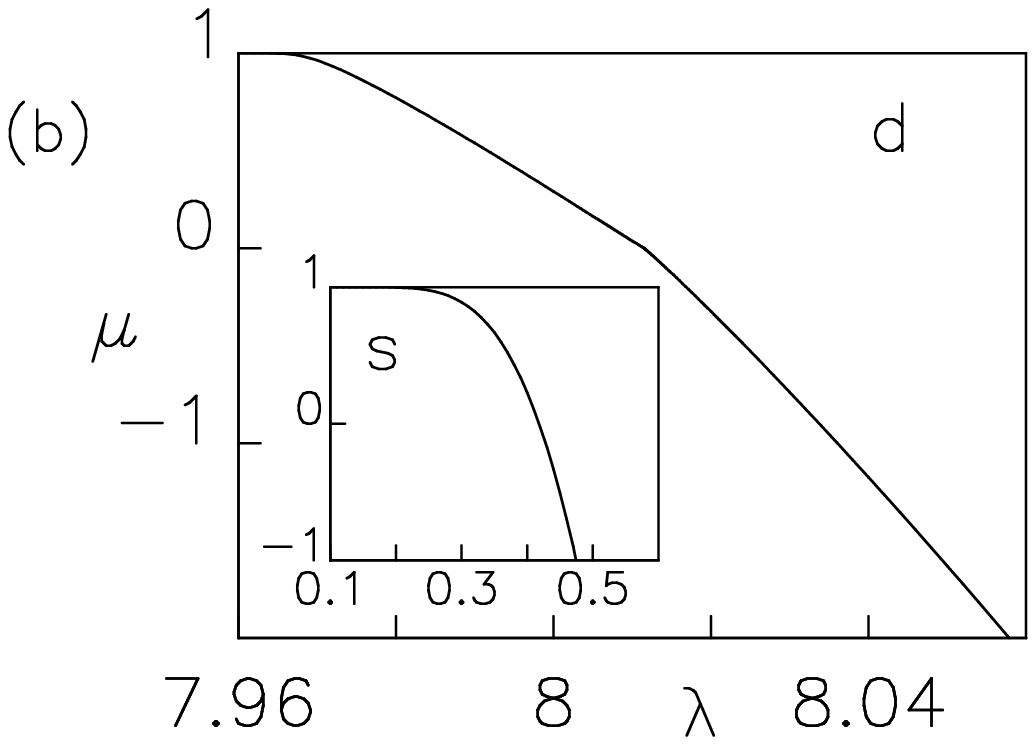}
\vskip -.7cm
    \caption{(a) The order parameter $\Delta_0$ and (b) the chemical 
potential $\mu$ as a function of coupling at fixed density $n = 1$ 
and $k_1=k_0=10$ for both s- and d-wave channels. 
}
\label{delta}
\end{center}
\end{figure}

Under these circumstances, 
quite generally it is not possible to find a separable potential 
in momentum space ${ V_{\bf k k^{\prime}} 
= -\lambda w^{*}({\bf k}) w({\bf k^\prime }) }$, nevertheless 
in the spirit of ref.~\cite{NSR} we choose to study a separable potential 
that contains most of the general features described above. 
In addition, we consider only singlet superconductivity, 
where the s-wave and the d-wave channels are studied separately.
For this purpose, we use the separable potential 
${ V_{\bf k k^\prime} 
= -\lambda_{\ell} w_{\ell}({\bf k}) w_{\ell}({\bf k^\prime }) }$. 
The interaction term ${ w_{\ell}({\bf k})}$ can be written as 
the product of two functions,
${
w_{\ell}({\bf k}) = h_{\ell} ({k}) g_{\ell} ({\bf \hat k}),
}$
where ${ h_{\ell} (k) = (k/k_1)^{\ell}
/{ \left[ 1 + (k/ k_0) \right]^{\ell + 1/2} } }$
controls the range of the interaction
and ${ g_{\ell} ({\bf \hat k}) = \cos (\ell \phi) }$ is
the angular dependence of the interaction. 
Here $k_0 \sim R_0^{-1}$ and  ${k_1}$ sets the scale at low momenta.
At zero temperature, we assume that pairing occurs with 
the same total momentum ${ {\bf q} = 0}$ only. This simplifying feature 
leads to the following saddle point and number equations,
\begin{equation}
\label{eqn:sad2}
{
{1 \over \lambda_{\ell} } 
= \sum_{\bf k} { |w_{\ell} ({\bf k})|^2 \over 2 E_{\ell} ({\bf k}) }
},
\end{equation}
\begin{equation}
\label{eqn:num}
{
n = 2 \sum_{\bf k} n_{\ell} ( {\bf k} ),
}
\end{equation}
where ${n_{\ell} ({\bf k})  = \left[ 1 - { (\epsilon_{\bf k} - \mu) 
/  E_{\ell} ({\bf k}) } \right]/2 } $ is the momentum distribution,
${ E_{\ell} ({\bf k}) 
= { \left ( (\epsilon_{\bf k} - \mu )^2 
+ |\Delta_{\ell} ({\bf k})|^2 \right )^{1/2},
}}$
is the single particle excitation energy,
and 
${
\Delta_{\ell} ({\bf k}) = \Delta_{0\ell} w_{\ell} ({\bf k}).
}$ is the order paremeter.
For a given interaction range $R_0 \sim k_0^{-1}$, 
the transition from the BCS limit (largely overlaping pairs) to the 
BEC limit of (weakly overlaping pairs) may occur either by changing the 
attraction strength $\lambda_{\ell}$ or the density $n$. In either
case, this evolution can be safely analyzed with the approximations used 
here provided that the system is {\it dilute} enough, i.e., $n \ll k_0^2$. 
This means that below a maximum density
$n_{max} \sim {k_F}_{max}^2$, the interaction range 
$R_0$ is much smaller than the interparticle spacing 
${k_F}^{-1}_{max}$, ${R_0 \ll {k_F}_{max}^{-1} }$,
or equivalently $k_0/{k_F}_{max} \gg 1$.
Thus we choose to scale all energies with respect
to the maximal Fermi energy ${\epsilon_F}_{max}$,
which fixes the maximum density
${n = n_{max} = 2 \rho {\epsilon_F}_{max}}$,
and all momenta with respect to
${k_F}_{max} = \sqrt{2m{\epsilon_F}_{max}}$. The coupling
constant is scaled with respect to the two-dimensional 
density of states $\rho$.
From now on we use this scaling. The numerical solutions for 
${ { \Delta}_{0\ell}}$ and 
${ { \mu}}$, when ${ k_1} = { k_0} = 10$  
are shown in Fig. \ref{delta},
for fixed density ${n} = 1$,
and changing ${ { \lambda}_{\ell} }$. 
Similar plots can also be made for fixed interaction
and varying density $n$.
\vskip -.7cm
\vskip -.4cm
In the BCS limit the amplitude of the order parameter ($\phi = 0$)
is given by
$$\Delta_{\ell}({ k_\mu}) 
\sim \exp 
\left[2({\lambda^{-1}_{0\ell}}({ k_\mu}) 
-  {\lambda^{-1}_\ell})/{h_\ell}^2({ k_\mu})\right] .$$
With our choice of $h_\ell({ k})$, 
${ \lambda_{0d}}({ k_\mu}) \simeq 8 + { \mu}/24{ \epsilon_1}
+ {\cal O} (\left [{ \mu}/{ \epsilon_1} \right ]^2 )$,
valid for ${ \mu}/{ \epsilon_1} \ll 1$, where 
$\epsilon_1 = {k_1}^2$.
The ratios between $\Delta_{\ell}(k_\mu)$,
and the critical temperature $T_{c\ell}$ satisfy
the usual relations $\Delta_{s}(k_\mu)/T_{cs} = 1.76$,
and $\Delta_{d}(k_\mu) / T_{cd} = 2.14$.
In Fig.~\ref{delta} $\Delta_{0d}$ and $\mu$ have 
a second order discontinuity as a function of $\lambda_d$.
This discontinuity occurs when $\mu = 0$ in both $\Delta_{0d}$ and $\mu$,
for varying interaction $\lambda_d$ or varying density $n$.
The line ${ {  \mu}  = 0}$ for a d-wave system
is very special as it will be seen in the following discussion of
spectroscopic quantities.

The first spectroscopic quantity to be analyzed 
is the single quasiparticle excitation spectrum 
$E_{\ell} ({\bf k})$.
Let us discuss first the s-wave case in the zero range interaction limit
${ k_0} \to \infty$.
For ${ \mu > 0}$ the excitation spectrum 
has an isotropic gap at $k=k_\mu$,
$E_g (k_\mu) = |\Delta_{s} ({k}_{\mu})|$. This gap 
is completely isotropic in the vicinity of ${ {k}_{\mu} }$. 
At the intermediate regime, when $\mu = 0$, 
the gap takes the value
${ E_g (0) = |\Delta_s (0)|}$, when the chemical potential 
becomes negative towards the BEC limit, the minimum of the energy gap 
remains at ${ {\bf k } = 0}$, 
${E_g (0) = {\left ( \mu^2 +  |\Delta_s (0)|^2 \right )^{1/2} } }$.
When ${k_0}$ is finite the position of the minimum gap changes, but
the excitation spectrum is always gapped.

\begin{figure}
\begin{center}
\epsfxsize=4.2cm
\epsfbox{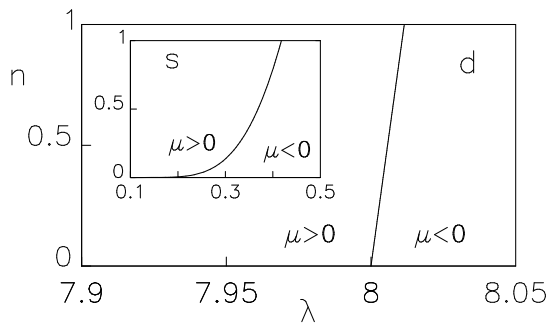}
\vskip -.6cm
    \caption{The line $\mu=0$ for both s- and d-wave order parameters
for $n=1$ and $k_1 = k_0 = 10$.}
\label{mu=0}
\end{center}
\end{figure}

In the d-wave case the situation is qualitatively different.
For ${ \mu > 0}$, including the BCS limit,  the excitation spectrum is gapless
at $k_\mu$ along the special directions ${ \phi = \pm \pi/4, \pm 3\pi/4}$,
near which the excitation spectrum disperses
linearly with momentum.
The energy gap at $ k = k_{\mu} $ and $\phi = 0$, 
$E_g (k_\mu) = |\Delta_{d} (k_{\mu}) | $
is a nonmonotonic function of $k_{\mu}$ for fixed
density, and thus a nonmonotonic function of ${\lambda_d}$.
The maximum $E_g (k_\mu)$ is reached at intermediate values of $\mu>0$.
At ${ \mu = 0}$, the minimal gap is ${ E_g (0) = |\Delta_d (0)| = 0}$, and 
occurs at the single point ${ {\bf k} = 0}$. 
In this case the excitation spectrum is 
${ E_{d} ({\bf k}) = { \left ( \epsilon_{\bf k}^2 
+ |\Delta_{d} ({\bf k})|^2 \right )^{1/2}} }$, which
behaves quadratically for small momenta at any given angle ${ \phi}$, 
since ${ \Delta_{d} ({\bf k}) \sim  k^2 \cos({2\phi}) }$ 
and ${ \epsilon_{\bf k} = k^2/2m}$.  
The shrinking of the energy gap to zero at ${\bf k} = 0$
is a consequence of the diminishing pairing interaction $h_d (k_\mu)$ 
for $\mu \to 0$. 
As soon as ${\mu < 0}$, including the BEC limit, a full gap in the excitation 
spectrum appears, but the minimal gap remains at ${ {\bf k} = 0}$ 
with value ${ E_g (0) = |\mu|}$ since ${ \Delta_{d} (0)} = 0$. 
Thus, the $\mu=0$ line separates a gapless d-wave superconductor ($\mu > 0$)
from a fully gapped d-wave superconductor ($\mu < 0$).
Fig.~\ref{mu=0} shows the lines where $\mu=0$ on the graph 
of $n$ vs. $\lambda_\ell$. 
Notice in Fig.~\ref{mu=0} that the low density limit of the s-wave system 
is always Bose-like, i.e., a two-body bound state appears at 
arbitrarily small $\lambda_s$. On the other hand, the d-wave system is  
qualitatively different: it is BCS-like for $\lambda_d < \lambda_{cd}$ 
and Bose-like for $\lambda_d > \lambda_{cd}$, where the critical
coupling $\lambda_c$ separating the two regimes  
is finite, i.e., the appearance of a two-body bound state 
in the d-wave case requires finite $\lambda_d$.
\vskip -1.8cm
\vskip -1.cm
The behavior of the excitation spectrum as a function of 
${ \mu}$ has important consequences on the momentum distribution
$n_{\ell} ({\bf k})$
at zero temperature. In the s-wave case the momentum distribution 
is isotropic in momentum space.
Here we discuss
briefly the behavior at low momenta for three different regimes:
${ \mu > 0}$, ${ \mu = 0}$, and ${ \mu < 0}$. Let us start with
${ \mu > 0}$. In the vicinity of ${ { k}_{\mu} }$ the momentum 
distribution is
${ n_s ({ k}_{\mu} + {\delta  k}) \simeq
\left[ 1 - 2 { k}_{\mu} \delta { k} /
\Delta_s ( k_\mu ) \right]/2 }$.
At low $k$ it behaves as
${n_s ({ k}) \simeq \left[ 1 
+ \gamma_p ( 1 + \alpha { k}/2{ k}_0 ) \right]/2 }$,
where
${\gamma_p = { \mu} / \sqrt{ { \mu}^2 + \Delta_{0s}^2 } }$, 
and
${\alpha = { \Delta}_s^2 / ( { \mu}^2 + \Delta_{0s}^2 ) }.$
When 
$\mu = 0$, 
the momentum distribution at small momenta is
${n_s (k) \simeq  \left( 1 - k^2 
/ \Delta_{0s} \right)/2 }$.
For negative ${ \mu}$,
${n_s ({ k}) = \left[ 1 - \gamma_n 
( 1 + \alpha { k}/2{ k_0} ) \right]/2 }$
for small $k$, with
${\gamma_n = |{ \mu}| / \sqrt{ { \mu}^2 + { \Delta}_{0s}^2 } }$.
Notice that $n_s(0)$ 
is a continuous function of ${  \mu}$.
In fact ${n_s ( { k} )}$ is a smooth function of 
${ \mu}$ for all momenta. This is not the case for a d-wave 
system, which shall be discussed next.
\vskip -0.8cm
\begin{figure}
\begin{center}
\epsfxsize=5cm
\hskip 2.5cm
\epsfbox{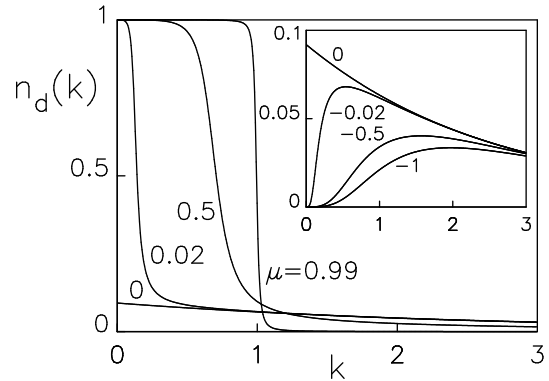}
\vskip -.8cm
    \caption{The momentum distribution of quasiparticles
for $\phi=0$, $n=1$, $k_1=k_0=10$,  and several values of 
$\mu$ for a d-wave order parameter.
The inset shows results for $\mu \leq 0$.}
\label{nk}
\end{center}
\end{figure}
\vskip -.4cm
The momentum distribution in
in the d-wave case is anisotropic, 
having the form ${ n_d ({ k}) = \left[ 1 - {\rm sgn} 
({ k}^2 -  \mu ) \right] }$ along the direction of the nodes 
(${\phi = \pm \pi/4, \pm 3 \pi/4}$). This behavior already signals 
discontinuity of $n_d(k)$ as a function of $\mu$ at $k=0$,
a suspicion further confirmed by analyzing the more interesting direction
${\phi = 0}$ and its equivalents ${\rm \phi =  \pm \pi/2, \pi}$.
Near $k_\mu$ the momentum distribution is
${n_d ({ k}_{\mu} + {\delta  k}) \simeq  
\left( 1 - 2 { k}_{\mu} \delta { k} /
{ \Delta}_d ( { k}_{\mu} )  \right)/2 }$.
On the other hand, the momentum 
\hbox{distribution behaves as}
${n_d ({ k}) \simeq 
1 - ( { \Delta}_{0d}^2/{ \mu}^2) ({ k}^4 / 4{ k}_1^4)
}
$
for small momenta. When $\mu=0$
the momentum distribution 
\hbox{at $k=0$ is ${n_d ({0}) \simeq  \left(1 - \kappa \right)/2 }$,}
where ${\kappa = ( 1 + \Delta_{0d}^2/{ k}_1^4 )^{-1/2} }$.
When ${ { \mu} < 0 }$, then
${ n_d ({ k}) \simeq  ( \Delta_{0d}^2/{ \mu}^2 ) 
({ k}^4 /4 { k}_1^4) }$
for small $k$. Notice the discontinuity of the momentum 
distribution at low $k$, when 
chemical potential crosses zero.
This discontinuity, which is illustrated in Fig.~\ref{nk},
coincides with the collapse of the four Dirac points 
to a single point at ${k_{\mu} = 0}$, and with the appearance 
of a full gap as soon as ${\mu < 0}$.

\begin{figure}
\begin{center}
\epsfxsize=4.5cm
\hskip 0cm
\epsfbox{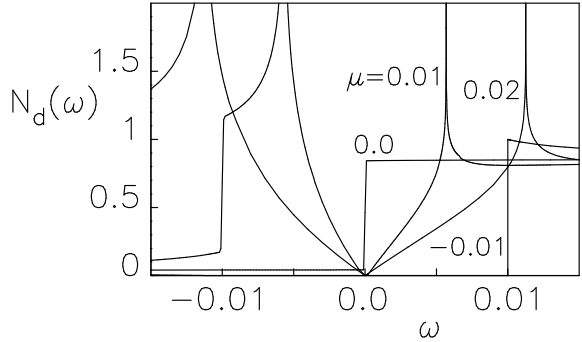}
\vskip -1.5cm
    \caption{Density of states for a d-wave order parameter near $\mu=0$,
for n = 1, $k_1=k_0=10$, and varying $\lambda_d$.
Notice the strong asymmetry (absence of quasiparticle-quasihole symmetry) 
between $\omega > 0$ and $\omega < 0$. }
\label{dos}
\end{center}
\end{figure}

The qualitative changes in $E_\ell ({\bf k})$ and $n_\ell ({\bf k})$, 
as a function of ${ \mu}$, affect substantially the quasiparticle 
density of states 
$N_{\ell} (\omega) = N_\ell^{(+)} (\omega) + N_\ell^{(-)} (\omega)$, 
where 
\begin{equation}
N_\ell^{(+)} (\omega) = 
{\left[2 \pi \right]^{-1}}
\int d^2{\bf k} \left[ 1 - n_{\ell} ({\bf k}) \right]
\delta (\omega - E_{\ell} ({\bf k}) ),
\end{equation}
corresponds to adding a quasiparticle, and
\begin{equation}
{ N_{\ell}^{(-)} (\omega) =
{\left[ 2 \pi \right]^{-1}}
\int d^2{\bf k}
n_{\ell} ({\bf k}) \delta (\omega + E_{\ell} ({\bf k}) ), }
\end{equation}
corresponds to removing a quasiparticle. 
In the s-wave case $N_{s} (\omega)$ is always zero at low frequencies, 
since the excitation spectrum is gapped for all ${\mu}$. 
On the other hand, 
${N_{d} (\omega)}$ changes discontinously at low frequencies,
from linear in ${\omega}$ for ${\mu > 0}$, where ${E_d ({\bf k})}$ is
linear in momentum close to the nodes, to a constant at
${ \mu = 0}$ (where $E_d ({\bf k}) \propto k^2$ at low $k$),
to zero for $\mu < 0$ (where $E_d ({\bf k})
\simeq  |\mu| + {\cal O}(k^2)$ for small $k$), as can be seen in
Fig.~\ref{dos}. 

\vskip -1.cm
\vskip -.4cm

Lastly, it is important to point out qualitative differences in thermodynamic
quantities, e.g. specific heat $C$ and spin susceptibility $\chi$, during the
evolution from BCS to BEC superconductivity at low temperatures.
The contributions from 
quasiparticles to $C$ and $\chi$ are exponentially small at low temperatures
in the s-wave case 
for all $\mu$, 
given that the excitation spectrum is always gapped. 
The situation is qualitatively different in the d-wave case 
where $C \propto T^2$, and $\chi \propto T$ for $\mu > 0$;
$ C \propto T $, and $\chi \propto {\rm const.}$ for $\mu = 0$; 
$C \propto T^{-1} \exp (-|\mu|/T )$, and $\chi \propto \exp (-|\mu|/T)$
for $\mu < 0$.

In summary we studied  
the low temperature evolution from BCS to BEC superconductivity for varying 
density and interaction strength in both s-wave and d-wave channels. 
In the s-wave case the excitation spectrum is always gapped, and the momentum
distribution is a continuous function of $\mu$.
However, in the d-wave case the excitation spectrum is gapless for $\mu >0$
and acquires a full gap for $\mu < 0$. Furthermore, the momentum 
distribution is discontinuous at low $k$, as $\mu$ crosses zero. 
As a result, the changes in spectroscopic and thermodynamic properties
near $\mu=0$ are dramatic at low temperatures.~\cite{lastfoot}
The line $\mu = 0$ in the
$n$ vs. $\lambda_d$ plane seems to correspond to a quantum critical line.

We are grateful to A.J. Leggett and E. Abrahams for discussions.
We would like to thank the Georgia Institute of Technology for 
financial support. Part of the numerical calculations were
performed on the Cray J916 at the Pozna\'n Supercomputer Center.

\vskip .2cm
\noindent
$^*$ Permanent address: Institute of Physics, A. Mickiewicz University,
Umultowska 85, 61-614 Pozna\'n, Poland.


\begin{references}
\bibitem{lastfoot} It is important to realize that the singular behavior of
spectroscopic and thermodynamic quantities 
(at low frequencies and temperatures) near $\mu = 0$
 in a d-wave superconductor 
is quite general and largely independent of details of the pairing potential.
\end{references}
\end{document}